\documentclass[10pt,conference]{IEEEtran}
\usepackage{graphicx, graphics,epsfig}
\usepackage{color}
\usepackage[cmex10]{amsmath}
\usepackage{amssymb}
\usepackage{url}
\usepackage{amsfonts}
\usepackage{latexsym}
\usepackage{tabularx}
\usepackage{stmaryrd}
\usepackage{bbm}
\usepackage{subfigure,dsfont,comment,colortbl,cite}
\usepackage{algorithm}
\usepackage{algorithmic}
\usepackage{multirow}

\newtheorem{thm}{\bf Theorem}
\newtheorem{prop}{\bf Proposition}

\title{\textbf{\fontsize{16pt}{16pt}\selectfont Optimal Distributed Broadcasting with Per-neighbor
Queues}}

\begin{document}
\maketitle

\begin{abstract}
Broadcasting systems such as P2P streaming systems represent important network applications
that support up to millions of online users. An efficient broadcasting
mechanism is at the core of the system design. Despite substantial efforts on developing efficient broadcasting algorithms, the following important question remains open: How to achieve the maximum broadcast rate in a distributed manner with each user maintaining information queues only for its direct neighbors? In this work, we first derive an innovative formulation
of the problem over acyclic overlay networks with arbitrary underlay capacity
constraints. Then, based on the formulation,
we develop a distributed algorithm to achieve the maximum broadcast
rate and every user only maintains one queue per-neighbor. Due to its lightweight nature, our
algorithm scales very well with the network size and remains robust against high system
dynamics. Finally, by conducting simulations we validate the optimality
of our algorithm under different network capacity models. Simulation results further indicate that
the convergence time of our algorithm grows linearly with the
network size, which suggests an interesting direction for future investigation.
\end{abstract}

\section{Introduction}

Information broadcasting is an increasingly important application for content
delivery and network control in the Internet.
For example, in P2P streaming systems such as PPLive
and UUSee, the streaming server continuously generates
streaming contents and disseminates them to all the participating
users.

We study the optimal broadcasting problem, {\em i.e.}, maximizing the
broadcast rate at which contents are received by all the users simultaneously.
This fundamental problem is at the core of many broadcasting
systems, attracting a substantial level of attention from both industry and academia.
A number of work in the literature design broadcasting
algorithms under various network models. We consider a general network model
that subsumes most known ones such as the edge-capacitated network model
and the node-capacitated network model, as illustrated in Fig. \ref{fig:overlay_network}.

A classic work on broadcasting is due to Edmonds \cite{all:article:edmonds73},
who provided a centralized algorithm for computing the maximum
broadcast rate. Edmonds' algorithm constructs a set of edge-disjoint
spanning trees, each rooted at the source and reaching every
user in the network. However it works only under the edge-capacitated
network model. Under the node-capacitated network model, Sengupta {\em et al.} \cite{sudipta2009lcclc}
formulated the problem as finding the set of spanning trees with maximum aggregate broadcast rate, subject
to node capacity constraints. They then solved this optimization problem approximately by adapting the
Garg-Konemann technique \cite{garkon}. These two \emph{centralized routing}
algorithms achieve the maximum broadcast rate, but both need an centralized entity to collect global network information, carry out the computation, and
coordinate the implementation of the solution, making them less adaptive to system dynamics due to
frequent reconstruction of trees.

Distributed solutions to the optimal broadcasting problem were then
provided by Ho and Viswanathan \cite{ho2009dynamic} and by Zhang {\em et al.} \cite{zhang2010icnp}, under the
edge-capacitated network model and the node-capacitated network model,
respectively. These solutions combine random network
coding \cite{ho2009dynamic}, \cite{chou2003practical} and back-pressure
based capacity scheduling \cite{Tassiulas92backpressure}. Although
the algorithms are distributed, each participating user needs to \emph{maintain one
queue for every other user in the entire network}, which stores contents intended for
that user. As the network size grows, the storage
and communication overhead introduced by maintaining and updating
all the queues at each user soon becomes prohibitive. Furthermore, upon user joins and
departures, the whole network needs to be informed to add or remove queues,
which is rather inefficient.

Massoulie {\em et al.}~\cite{massoulie2007rdb} proposed a simple distributed
algorithm for both edge-capacitated and node-capacitated
networks. Every user only needs to maintain one queue for each of
its neighbors, which stores contents innovative to the corresponding
neighbor. It is proved to support any feasible broadcast rate for
arbitrary edge-capacitated networks and full-mesh node-capacity networks.
But the algorithm fails when the capacity bottleneck is on the underlay links such as
the example in Fig. \ref{fig:overlay_network}. Also it works only if a feasible target broadcast rate is given. It
\emph{cannot approach the maximum broadcast rate} adaptively.

Besides, several papers \cite{zhang2010icnp}, \cite{streaming_capacity.icdcs10}, \cite{lin.infocom13.capacity} studied the broadcasting capacity when user can select its direct neighbors from the network, and designed algorithms to approach it.

Despite all the existing results, the problem remains very challenging if we wish to achieve the maximum
broadcast rate in a distributed fashion, while \emph{maintaining only per-neighbor information queue at each node}.
There is no existing problem formulation
to help design such an algorithm. The problem becomes even harder when one considers
adapting the algorithm to a general network model. Under our network model, the algorithm
needs to learn underly capacity bottlenecks and adjust the overlay link rate accordingly.

In this work, we aim at designing such a broadcasting
algorithm and make the following contributions:
\begin{enumerate}
\item We consider the model that the overlay network
is acyclic but underlay capacity bottlenecks can be anywhere, which subsumes most known ones.
Under this model, we formulate
the optimal broadcasting problem in an innovative way. We show that our formulation is simple
yet effective to characterize any feasible broadcast rate. Most importantly,
this formulation leads us to efficient algorithm design.
\item Based on our formulation, we are the first to design a distributed
algorithm to solve the optimal broadcasting problem, which requires
only per-neighbor information maintained at each node. The proposed
algorithm scales very well and is robust to high levels of network
dynamics.
\end{enumerate}
We summarize the existing results and our result in Table \ref{tab:comparison}.

\begin{table*}[htbp]
\begin{centering}
\begin{tabular}{|c|c|c|>{\centering}p{1.5cm}|>{\centering}p{1.5cm}|>{\centering}p{1.5cm}|>{\centering}p{2cm}|c|>{\centering}p{2.8cm}|}
\hline
\multirow{2}{*}{{\scriptsize References}} & \multicolumn{2}{c|}{{\scriptsize Network Topology}} & \multicolumn{3}{c|}{{\scriptsize Network Capacity Bottleneck}} & \multirow{2}{2cm}{{\scriptsize Can automatically learn the maximum broadcast rate?}} & \multirow{2}{1.2cm}{{\scriptsize Distributed?}} & \multirow{2}{2.8cm}{{\scriptsize Number of Queues Maintained at Each User}}\tabularnewline
\cline{2-6}
 & {\scriptsize Acyclic?} & {\scriptsize Cyclic/Full-Mesh?} & {\scriptsize Edge Capacitated?} & {\scriptsize Node Capacitated?} & {\scriptsize Arbitrary Bottleneck?} &  &  & \tabularnewline
\hline
{\scriptsize \cite{all:article:edmonds73}} & $\times$ & $\surd/\times$ & $\surd$ & $\times$ & $\times$ & $\surd$ & $\times$ & {\scriptsize $O(1)$}\tabularnewline
\hline
{\scriptsize \cite{sudipta2009lcclc}, \cite{streaming_capacity.icdcs10}} & $\times$ & $\surd/\times$ & $\times$ & $\surd$ & $\times$ & $\surd$ & $\times$ & {\scriptsize $O(1)$}\tabularnewline
\hline
{\scriptsize \cite{ho2009dynamic}} & $\times$ & $\surd/\times$ & $\surd$ & $\times$ & $\times$ & $\surd$ & $\surd$ & {\scriptsize $O(\mbox{number of total users})$}\tabularnewline
\hline
{\scriptsize \cite{zhang2010icnp}} & $\times$ & $\surd/\times$ & $\times$ & $\surd$ & $\times$ & $\surd$ & $\surd$ & {\scriptsize $O(\mbox{number of total users})$}\tabularnewline
\hline
{\scriptsize \cite{lin.infocom13.capacity}} & $\times$ & $\surd/\times$ & $\times$ & $\surd$ & $\times$ & $\times$ & $\surd$ & {\scriptsize $O(\log (\mbox{number of total users}))$}\tabularnewline
\hline
{\scriptsize \cite{massoulie2007rdb}} & $\times$ & $\surd/\times$ & $\surd$ & $\times$ & $\times$ & $\times$ & $\surd$ & {\scriptsize $O(\mbox{number of neighbors})$}\tabularnewline
\hline
{\scriptsize \cite{massoulie2007rdb}} & $\times$ & $\surd/\surd$ & $\times$ & $\surd$ & $\times$ & $\times$ & $\surd$ & {\scriptsize $O(\mbox{number of neighbors})$}\tabularnewline
\hline
{\scriptsize This work} & $\surd$ & $\times/\times$ & $\surd$ & $\surd$ & $\surd$ & $\surd$ & $\surd$ & {\scriptsize $O(\mbox{number of neighbors})$}\tabularnewline
\hline
\end{tabular}
\par\end{centering}

\caption{Summary and comparison of previous work and this work for the optimal
broadcasting problem.\label{tab:comparison}}
\end{table*}

\section{Problem Setting and Notations}

A network is modeled as a directed graph $\mathcal{G}=(\mathcal{N},\mathcal{L})$,
where $\mathcal{N}$ is the set of physical nodes including all
the broadcasting participants and other intermediate nodes such as
routers, and $\mathcal{L}$ is the set of all physical links.

\begin{figure}
\begin{centering}
\includegraphics[scale=0.09]{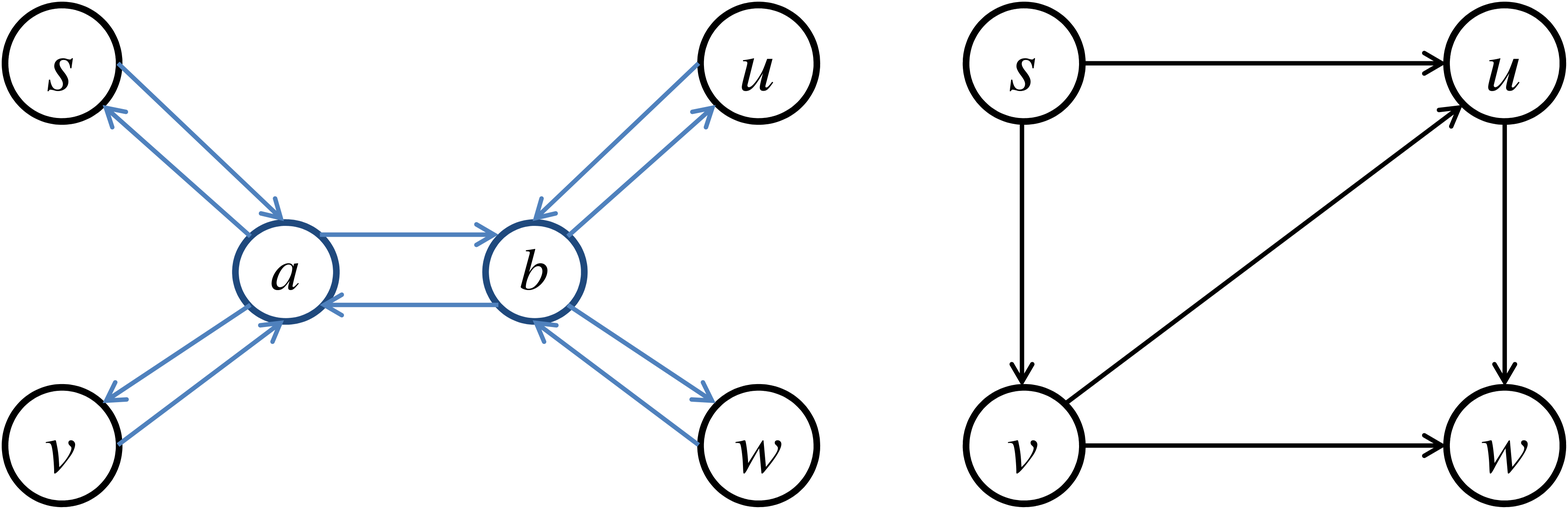}
\par\end{centering}
\caption{Network Model: The graph on the right represents
an overlay network built upon the network on the left.
The broadcasting system works in the overlay network.
Nodes $a$ and $b$ are routers; nodes $s$,$u$,$v$ and $w$
denote broadcasting participants. An edge is a physical
link on the left, and is a multi-hop TCP/UDP connect on the right. For example, the overlay
link $(v,u)$ traverses the physical path $(v,a)\rightarrow(a,b)\rightarrow(b,u)$.
The capacity constraint can be anywhere --- for example, on the overlay links ({\em edge-capacitated}) or
on the broadcasting nodes ({\em node-capacitated}).
}
\label{fig:overlay_network}
\end{figure}

Consider a single source broadcasting system deployed in an overlay network
$G=(V,E)$ built upon $\mathcal{G}$. We assume that $G$
is acyclic. The node set $V\subseteq \mathcal{N}$ represents the set of
all the broadcasting participants. The edge set $E$ represents the
set of \emph{overlay} links connecting the nodes $V$. An \emph{overlay}
link is a \emph{TCP} or \emph{UDP} connection, which may traverse
multiple physical links of $\mathcal{L}$. A simple example
is shown in Fig. \ref{fig:overlay_network}. Let $s\in V$
denote the source and $R=V-\{s\}$ denote the set of receivers. Let
\[
in(v)=\{w\in V|\,(w,v)\in E\}
\]
be the set of incoming neighbors of node $v$, from which node $v$
may receive contents. Let
\[
out(v)=\{u\in V|\,(v,u)\in E\}
\]
be the set of outgoing neighbors of node $v$, to which node $v$
can transmit contents. The source $s$ generates contents continuously
at rate $z$. Each receiver attempts to collect all the contents from
its incoming neighbors.

Let $r_{vu}$ be the content transmission rate over link $(v,u)\in E$.
The network capacity constraints are expressed as follows:
\begin{equation}
A\cdot\boldsymbol{r}\leq\boldsymbol{C},\label{eq:capacity_constraint}
\end{equation}
where the column vector $\boldsymbol{r}=\{r_{vu},\,(v,u)\in E\}$
denotes the link rate set, the column vector $\boldsymbol{C}$ denotes
all the capacity bottlenecks in the network and the matrix $A$ reflects
how links share the network capacities. \emph{Note that our network
capacity model is rather general and the inequality (\ref{eq:capacity_constraint})
subsumes scenarios where the capacity bottleneck can be anywhere
in the network.} For instance, for a P2P broadcasting
system where the capacity constraint is generally assumed to be on
the nodes, {\em i.e.}, any node $v$'s aggregate outgoing rate $\sum_{u\in out(v)}r_{vu}$
is upper bounded by the node $v$'s upload capacity $C_{v}$. Then
we can write $\boldsymbol{C}=\{C_{v},\, v\in V\}$ and $A=\{a_{v,e},\, v\in V,e\in E\}$
where
\[
a_{v,e}=\begin{cases}
1 & \mbox{if the head of }e\mbox{ is }v,\\
0 & \mbox{otherwise}.
\end{cases}
\]
For example, in Fig. \ref{fig:overlay_network}, for node $v$, we
have $r_{vu}+r_{vw}\leq C_{v}$. While if the capacity constraint
is on the physical links, for example, in wireless communication systems,
we can write $\boldsymbol{C}=\{C_{l},\, l\in\mathcal{L}\}$ and $A=\{a_{l,e},\, l\in\mathcal{L},e\in E\}$
where $C_{l}$ is the link capacity and
\[
a_{l,e}=\begin{cases}
1 & \mbox{if overlay link }e\mbox{ passes physical link }l,\\
0 & \mbox{otherwise}.
\end{cases}
\]
For example, in Fig. \ref{fig:overlay_network}, for the physical link
$(a,b)$, we have $r_{su}+r_{vu}+r_{vw}\leq C_{ab}$. For the convenience
of illustration, we use $e$ and $(v,u)$ exchangeably to denote one
edge in $E$ afterwards.

\section{Problem Statement}

Before presenting our problem, we first characterize the maximum
broadcast rate $B$, {\em i.e.}, the highest rate at which every node in
the system can receive the contents simultaneously.

Given an $\boldsymbol{r}=\{r_{vu},\,(v,u)\in E\}$ satisfying (\ref{eq:capacity_constraint}),
for any $v\in R$, we let $\rho_{s,v}(\boldsymbol{r})$ be the minimum
$s-v$ cut capacity. $\rho_{s,v}(\boldsymbol{r})$ can be expressed
as follows
\begin{eqnarray}
\rho_{s,v}(\boldsymbol{r})= & \underset{U,\bar{U}}{\min} & \sum_{(u,w)\in E,u\in U,w\in\bar{U}}r_{uw}\label{eq:min_cut_def}\\
 & \mbox{s.t.} & s\in U,\, v\in\bar{U},\nonumber \\
 &  & U\cup\bar{U}=V,\nonumber \\
 &  & U\cap\bar{U}=\phi,\nonumber
\end{eqnarray}
where the three constraints define that $(U,\bar{U})$ is a $s-v$
cut, and the objective is the cut capacity. By the well-known max-flow
min-cut theorem, $B$ is equal to the value of the following problem
\begin{align}
\max_{\boldsymbol{r}} \min_{v\in R} & \,\,\rho_{s,v}(\boldsymbol{r})\label{eq:cut_based_pf}\\
\mbox{s.t.} & \,\,A\cdot\boldsymbol{r}\leq\boldsymbol{C}.\nonumber
\end{align}
In other words, the maximum broadcast rate is equal to the maximum
of the minimum source-receiver cut capacities across all feasible
link rate allocations.

We wish to design a distributed algorithm to achieve the maximum
broadcast rate $B$, and at the same time satisfy the requirement
$\boldsymbol{Neighbor-only}$: \emph{only neighbor-regarded information is
maintained at each node}. For example, every node may keep each neighbor's
current state (e.g., hitherto received contents) and update it periodically
in the algorithm. Such information is generally used to guide node's
behavior. The reason that only neighbor-regarded information is
maintained is to guarantee good scalability and robustness. With only neighbor-regarded information, the
storage and communication overhead is lightweight
even in large-scale systems. Upon node joins/departures,
we only need to notify the corresponding neighbors. Consequently, even
if the system is highly dynamic, the algorithm can still adapt in
an agile and efficient way. We will further explain the requirement $\boldsymbol{Neighbor-only}$
and its benefits in Section \ref{sub:broadcasting_algorithm} when
we describe our broadcasting algorithm.

\section{Problem Formulation and Justification}

\subsection{Problem Formulation}

The optimal broadcasting problem can be formulated in different ways.
Besides (\ref{eq:cut_based_pf}), the information-flow
based formulation is most common \cite{ho2009dynamic}, \cite{zhang2010icnp}.
However, none of the algorithms based on existing formulations can
satisfy the strict requirement of $\boldsymbol{Neighbor-only}$. Our
new formulation below enables us to design a broadcasting algorithm
that is throughput optimal and distributed, and most importantly, satisfies
the $\boldsymbol{Neighbor-only}$ condition.

We formulate the broadcast rate maximization problem as follows:
\begin{align}
\max_{z\geq0,\boldsymbol{r}\geq0} & U(z)\label{eq:mp}\\
\mbox{s.t.} & \sum_{u\in in(v)}r_{uv}\geq\sum_{p\in in(w)}r_{pw}+z\boldsymbol{1}_{w=s},\forall v\in R,w\in in(v),\label{eq:mp_constraint1}\\
 & A\cdot\boldsymbol{r}\leq\boldsymbol{C},\label{eq:mp_constraint2}
\end{align}
where $z$ is the rate at which the source $s$ generates contents,
$U(\cdot)$ is a differentiable strictly concave increasing utility
function, and $r_{uv}$ is the allocated rate over the overlay link $(u,v)$ at
which node $u$ can transmit contents to node $v$. Constraint
(\ref{eq:mp_constraint1}) states that a node's total receiving
rate should be no less than the total receiving rate of each of its
incoming neighbors. The network capacity constraint is captured
by constraint (\ref{eq:mp_constraint2}).

Constraint (\ref{eq:mp_constraint1}) differentiates our problem
formulation fundamentally from existing ones in the literature. Prior to
this work, the information-flow based formulation is mostly used to help
design broadcasting algorithms. In the information-flow based formulation,
instead of (\ref{eq:mp_constraint1}), there is a flow-balancing
constraint for \textit{every node pair}. In our formulation, however,
(\ref{eq:mp_constraint1}) involves \textit{only node-neighbor pairs}.
This unique feature guarantees that, in our broadcasting algorithm
based on this formulation, every node interacts just with its neighbors
and only neighbor-regarded information is needed.

Now we explain roughly why the constraint (\ref{eq:mp_constraint1})
makes sense. It is known that, for any acyclic graph, we can
always divide the nodes into different ordered layers such that the
source is exclusively in the lowest layer and every node stays in
lower layer than any of its outgoing neighbors. In other words, nodes
can only receive contents from those in lower layers. Let's consider
nodes layer by layer from bottom to top. First, nodes in the second
lowest layer can receive contents only from the source. The constraint
(\ref{eq:mp_constraint1}) makes sure that these nodes can obtain
whatever the source has, since the receiving rate is no less than
$z$. Next, nodes in the third lowest layer can receive the contents
from either the source or nodes in the second lowest layer. Due to
the similar reason, the constraint (\ref{eq:mp_constraint1}) guarantees
these nodes can also get all the source has. Then we can continue
the same argument until the highest layer. So essentially the constraint
(\ref{eq:mp_constraint1}) is to make every receiver receive all the
contents generated by the source.

Next, we formally show that the simple constraint (\ref{eq:mp_constraint1})
along with (\ref{eq:mp_constraint2}) give us the feasible region
of the broadcast rate for any acyclic graph.

\subsection{Formulation Justification}

In this subsection, we first establish a known property about the
acyclic graph $G$: nodes can be indexed so that every node's index
is smaller than that of any outgoing neighbor. Then, based on this
property, we show that constraints (\ref{eq:mp_constraint1}), (\ref{eq:mp_constraint2})
are sufficient and necessary conditions for any feasible broadcast
rate $z\leq B$.

The topological ordering of acyclic graphs is a known result. It is presented in the following proposition for the sake of completeness.
\begin{prop}
All nodes in $G$ can be sequentially indexed such that, for any
node $v\in R$, the index of any $u\in in(v)$ is smaller than $v$'s
index.\label{prop:acyclic_property}
\end{prop}

We say a broadcast rate $z$ is feasible if and only if $z\leq B$. Based
on Proposition \ref{prop:acyclic_property}, we show in the following
theorem that constraints (\ref{eq:mp_constraint1}), (\ref{eq:mp_constraint2})
are sufficient and necessary conditions for any feasible broadcast
rate under the acyclic directed graph. To show necessity, we prove
that, for any feasible $z$, we can find a $\boldsymbol{r}$ that
supports $z$ and at the same time satisfies constraints (\ref{eq:mp_constraint1}), (\ref{eq:mp_constraint2}).
On the other hand, for any $s-v$ cut in $G$, we can always find
a node such that the cut capacity is no less than the node's total
receiving rate. Constraint (\ref{eq:mp_constraint1}) guarantees
that every node's receiving rate is larger than or equal to the broadcast
rate $z$. So the broadcast rate $z$ is not larger than any cut capacity,
and thus feasible by the definition of $B$ in (\ref{eq:cut_based_pf}).
\begin{thm}
If $G$ is acyclic, any $z$ such that there exists a $\boldsymbol{r}=\{r_{e},e\in E\}$
satisfying the constraints (\ref{eq:mp_constraint1}), (\ref{eq:mp_constraint2})
is a feasible broadcast rate. On the other hand, for any feasible
broadcast rate $z$, there exists one $\boldsymbol{r}=\{r_{e},e\in E\}$
satisfying the constraints (\ref{eq:mp_constraint1}), (\ref{eq:mp_constraint2}).
\label{thm:sufficient_necessary}
\end{thm}
\begin{IEEEproof}
First, for any broadcast rate $z$ such that there exists one $\boldsymbol{r}=\{r_{e},e\in E\}$
satisfying the constraints (\ref{eq:mp_constraint1}), (\ref{eq:mp_constraint2}),
we show that $z\leq B$.

Consider a new graph $G^{'}=(V,E)$, i.e., the same as the old one.
For each edge $e\in E$, the
edge capacity is $r_{e}$. $B^{'}$ is the maximum broadcast rate
that can be achieved under $G^{'}$. We should have $B^{'}\leq B$.
Like $G$, $G^{'}$ is also acyclic. By Proposition \ref{prop:acyclic_property},
we can index all the nodes in $G^{'}$ such that every node can only
receive content from those with smaller indexes. Denote one of such
index sets by $\{I_{v},v\in V\}$. Then we have for $\forall v\in V$,
if $(u,v)\in E$, then $I_{u}<I_{v}$.

Next we prove the following inequality
\[
B^{'}=\min_{v\in R}\rho_{s,v}(\boldsymbol{r})\geq z,
\]
where $\rho_{s,v}(\boldsymbol{r})$ is as defined in (\ref{eq:min_cut_def})
the minimum $s-v$ cut capacity in $G^{'}$. Consider any $s-v$ cut
$(U,\bar{U})$ and its cut capacity is $\sum_{u\in U,w\in\bar{U}}r_{(u,w)}$.
It can be shown that there exists a node $w^{*}\in\bar{U}$ such that
for any other $w\in\bar{U}$ we have $I_{w^{*}}<I_{w}$. This means
given such node $w^{*}$ the node set $\{u|(u,w^{*})\in E\}$ is a
subset of $U$. Because $\sum_{u\in U:(u,w^{*})\in E}r_{(u,w^{*})}\geq z$
which is guaranteed by the constraint (\ref{eq:mp_constraint1}),
we have the cut capacity $\sum_{u\in U,w\in\bar{U}}r_{(u,w)}\geq z$.
Since this inequality is true for any $s-v$ cut, we prove that $z$
is no larger than the minimum of the minimum $s-v$ cut capacities.
Since $B^{'}\leq B$, $z$ is feasible for $G$.

Second, for any feasible broadcast rate $z$, Edmonds established
in \cite{all:article:edmonds73} that $z$ can be achieved by packing
spanning trees. Let $\mathcal{T}$ be the set of all spanning trees
in $G$. For a feasible $z$, there exists a $\boldsymbol{\lambda}=\{\lambda_{T},T\in\mathcal{T}\}$
such that
\[
\sum_{T\in\mathcal{T}}\lambda_{T}=z,
\]
 and
\[
A\cdot\boldsymbol{r}\leq\boldsymbol{C},
\]
where $\boldsymbol{r}=\{r_{e}|r_{e}=\sum_{T\in\mathcal{T}:e\in T}\lambda_{T}\}$.

Now we check that $\boldsymbol{r}$ satisfies all the constraints
(\ref{eq:mp_constraint1}), (\ref{eq:mp_constraint2}). The constraint
(\ref{eq:mp_constraint2}) is satisfied trivially. In any tree $T\in\mathcal{T}$,
every node's receiving rate is identical which is equal to $\lambda_{T}$.
So every node's total receiving rate is same as $\sum_{T\in\mathcal{T}}\lambda_{T}$.
Then the constraint (\ref{eq:mp_constraint1}) gets satisfied.
\end{IEEEproof}

We can get from the above proof that as long as we make sure every
node's receiving rate is larger than or equal to the broadcast rate
$z$, then every cut capacity is also guaranteed larger than or equal
to $z$. In other words, in the acyclic graph, all {}``critical''
cuts are contained in the cut set $\{(V\backslash v,\, v)|\, v\in R\}$,
whose capacity is equal to one specific node's receiving rate. {}``Critical''
means the capacity of any other cut is no less than that of at least
one cut in $\{(V\backslash v,\, v)|\, v\in R\}$. So the constraint
(\ref{eq:mp_constraint1}) can be understood as follows: the broadcast
rate $z$ should be no more than the capacity of any {}``critical''
cut. For any acyclic graph, we can characterize the feasible broadcast
rate without considering all the cuts%
\footnote{The Theorem \ref{thm:sufficient_necessary} can be extended to graphs
which satisfy that all {}``critical'' cuts belong to $\{(V\backslash v,\, v)|\, v\in R\}$%
}.

\section{Algorithm Design}
Based on our problem formulation, we now apply the classic
Lagrangian decomposition approach to design a broadcasting algorithm,
which is throughput optimal, distributed and maintains only neighbor-regarded
information at each node.

\subsection{Lagrangian Decomposition}

By relaxing constraint (\ref{eq:mp_constraint1}), we obtain the
following partial Lagrangian:
{\small
\begin{align*}
 & L(z,\boldsymbol{\theta},\boldsymbol{r})\\
= & U(z)+\sum_{v\in R}\sum_{w\in in(v)}\theta_{v,w}\left(\sum_{u\in in(v)}r_{uv}-\sum_{p\in in(w)}r_{pw}-z\mathbf{1}_{w=s}\right),
\end{align*}
}where $\theta_{v,w}$ is the Lagrange multiplier. 
%

Since the Slater constraint qualification conditions hold for the
problem (\ref{eq:mp}) \cite{Boyd04}, strong duality holds. Thus
problem (\ref{eq:mp}) can be solved by finding the saddle points
of $L(z,\boldsymbol{\theta},\boldsymbol{r})$, through solving the following
problem in $z,\boldsymbol{\theta},\boldsymbol{r}$:
\begin{align*}
\min_{\boldsymbol{\theta}\geq0} & \left(\max_{z\geq0}\left[U(z)-\sum_{v\in R:s\in in(v)}\theta_{v,s}z\right]+\right.\\
 & \left.\max_{\boldsymbol{r}\geq0}\sum_{v\in V}\sum_{u\in out(v)}r_{vu}\left[\sum_{w\in in(u)}\theta_{u,w}-\sum_{w\in out(u)}\theta_{w,u}\right]\right)\\
\mbox{s.t.} & \,A\cdot\boldsymbol{r}\leq\boldsymbol{C}.
\end{align*}

Given $\boldsymbol{\theta}$, we first solve the following capacity
scheduling subproblem in $\boldsymbol{r}$:
\begin{align*}
\mathbf{SSP}:\max_{\boldsymbol{r}\geq0} & \sum_{v\in V}\sum_{u\in out(v)}r_{vu}\left[\sum_{w\in in(u)}\theta_{u,w}-\sum_{w\in out(u)}\theta_{w,u}\right]\\
\mbox{s.t.} & \,A\cdot\boldsymbol{r}\leq\boldsymbol{C}.
\end{align*}

We can exploit the specific structure of the above linear program to solve
it in a distributed fashion. In particular, for any two neighboring
nodes $u$ and $v$, we define the back-pressure from $u$
to $v$ as
\begin{equation}
P_{vu}=\left[\sum_{w\in in(u)}\theta_{u,w}-\sum_{w\in out(u)}\theta_{w,u}\right],\;\forall(v,u)\in E.
\end{equation}

Under different scenarios of network capacity constraints,
we solve this subproblem correspondingly, and obtain the optimal solution
$\boldsymbol{r}^{*}$.

If capacity constraints are applied on the nodes:
\[
\sum_{u\in out(v)}r_{vu}\leq C_{v},\,\forall v\in V,
\]
for any $v\in V$, let $u^{*}(v)=\arg\max_{u\in out(v)}P_{vu}$, then
the solution to $\mathbf{SSP}$ is as follows: for any $v\in V$,
\begin{equation}
r_{vu}^{*}=\begin{cases}
C_{v}, & \mbox{if }u\in out(v),\, u=u^{*}(v)\mbox{ and }P_{vu}>0,\\
0, & \mbox{otherwise.}
\end{cases}\label{eq:node_sol}
\end{equation}

If capacity constraints are applied on the physical links, recall that they
can be written as:
\[
\sum_{e\in E}a_{l,e}r_{e}\leq C_{l},\,\forall l\in\mathcal{L},
\]
where $\mathcal{L}$ is the set of all physical links and
\[
a_{l,e}=\begin{cases}
1 & \mbox{if overlay link }e\mbox{ uses physical link }l,\\
0 & \mbox{otherwise}.
\end{cases}
\]
We can solve $\mathbf{SSP}$ by the following primal-dual algorithm
which will converge to $\boldsymbol{r}^{*}$
\begin{equation}
\begin{cases}
\dot{r}_{e}=\beta_{e}\left[P_{e}-\sum_{l\in\mathcal{L}}a_{l,e}\lambda_{l}\right]_{r_{e}}^{+},\,\forall e\in E,\\
\dot{\lambda}_{l}=\sigma_{l}\left[\sum_{e\in E}a_{l,e}r_{e}-C_{l}\right]_{\lambda_{l}}^{+},\,\forall l\in\mathcal{L},
\end{cases}\label{eq:underly_sol}
\end{equation}
where $\beta_{e}$ and $\sigma_{l}$ are positive constants, and function
\[
[b]_{a}^{+}=\begin{cases}
\max(0,b) & a\le0\\
b & a>0.
\end{cases}
\]

Given the $\mathbf{SSP}$'s solution $r_{vu}^{*}$ as (\ref{eq:node_sol})
or (\ref{eq:underly_sol}), we use the following distributed primal-dual
algorithm to solve the sub-problems in $z,\boldsymbol{\theta}$ simultaneously:
\begin{equation}
\begin{cases}
\dot{z}=\alpha\left[U^{'}(z)-\sum_{v\in V:s\in in(v)}\theta_{v,s}\right]_{z}^{+},\\
\dot{\theta}_{v,u}=\gamma_{v,u}\left[\sum_{p\in in(u)}r_{pu}^{*}+z\mathbf{1}_{u=s}\right.\\
\left.\qquad\qquad\quad-\sum_{w\in in(v)}r_{wv}^{*}\right]_{\theta_{v,u}}^{+},\,\forall v\in R,u\in in(v),
\end{cases}\label{eq:primal_dual}
\end{equation}
where $\alpha$ and $\gamma_{v,u}$ are positive constants.

\subsection{Broadcasting Algorithm\label{sub:broadcasting_algorithm}}

Our broadcasting algorithm works as follows: Every node $v$ maintains
one {}``queue'' $\theta_{v,u}$ for each incoming neighbor $u\in in(v)$.
In each time slot,
\begin{itemize}
\item \textbf{Primal-dual Update}: According to (\ref{eq:primal_dual}),
the source $s$ updates the broadcast rate $z$, and each node $v$
collects information from $u\in in(v)$ and updates the queue $\theta_{v,u}$.
\item \textbf{Capacity Scheduling}: Node $v$ decides the link transmission
rate $r_{vu}$ for all $u\in out(v)$ according to (\ref{eq:node_sol})
or (\ref{eq:underly_sol}).
\item \textbf{Content Scheduling}: Given $\boldsymbol{r}$, every node $v$
coordinates with its incoming neighbor set $\{u|\, u\in in(v),\, r_{uv}>0\}$,
and decides what to receive from each of them in order to obtain as
many innovative contents as possible. Then every node $u$ sends out
specific contents to each outgoing neighbor $v\in out(u)$ at rate
$r_{uv}$.
\end{itemize}

\noindent\textbf{Remark}: We can also adopt random network
coding to help content scheduling \cite{chou2003practical}, \cite{all:article:HMKKESL04}.
In each time slot, every node $u$ randomly encodes all the received
contents and sends out the coded contents to each $v\in out(u)$ at
rate $r_{uv}$. This way, there is no need for coordination between
each node-neighbor pair. However, network coding will introduce communication overhead
for carrying coding coefficients and computation complexity for encoding
and decoding.

We have the following observations for our broadcasting algorithm:
\begin{itemize}
\item The Lagrangian variable $\theta_{v,u}$ measures the buffer size difference
between node $v$ and its incoming neighbor $u$, which can be calculated
at node $v$ by collecting information from each of its neighbors.
\item With the above understanding on $\boldsymbol{\theta}$, the terms
in $P_{vu}$ can be understood as follows. The term $\sum_{w\in in(u)}\theta_{u,w}$
measures the aggregate deficit in received content amount between
node $u$ and all its incoming neighbors. The larger this term is,
the more desperate peer $u$ wants to receive contents. Similarly,
the term $\sum_{w\in out(u)}\theta_{w,u}$ measures the aggregate
surplus in received content amount between node $u$ and all its outgoing
neighbors.
\item The algorithm requires nodes to exchange information only with its
one-hop neighbors, and can be implemented in a distributed manner.
At each node, the maintained information is $\theta_{v,u}$ regarding
only incoming neighbors. Hence our algorithm satisfies the $\boldsymbol{Neighbor-only}$ condition.
The number of {}``queues'' every node
needs to maintain and update is just equal to half of the size of its
neighbors. As a result, the storage and communication cost of maintaining
and updating jobs at each node is quite limited even for large-scale
systems. When there is a node departing, that node just needs to notify
its outgoing neighbors, who eliminate the
{}``queue'' regarding the leaving node. When there is a new node joining,
that node first establishes one {}``queue'' for each incoming neighbor,
and then notifies its outgoing neighbors. Each outgoing neighbor adds
one {}``queue'' regarding the joining node. The adjustment overhead
incurred by network dynamic is only proportional to the neighbor size
and thus very lightweight. Our algorithm, therefore, scales very well
and is very robust to network dynamics.
\end{itemize}

Next, we show that our broadcasting algorithm is throughput optimal.
This means that our algorithm can guarantee high performance,
for example, high-quality video support, in practical systems.

To show this, we proceed in two steps. First, we show that the joint
primal-dual algorithm and capacity scheduling can converge to the
optimal solution $z^{*},\,\boldsymbol{r}^{*}$of (\ref{eq:mp}). Then,
under rate allocation $\boldsymbol{r}^{*}$, our content scheduling
strategy can guarantee that every node can receive innovative contents
at rate $z^{*}$. By Theorem \ref{thm:sufficient_necessary}, we get
$z^{*}=B$ which is the maximum broadcast rate. Overall, we have the
following theorem.
\begin{thm}\label{thm:alg_optimality}
Our broadcasting algorithm is throughput optimal.\end{thm}
\begin{IEEEproof}
By standard Lyapunov arguments, we can show that the joint primal-dual
algorithm and capacity scheduling can converge to the optimal solution
$z^{*},\,\boldsymbol{r}^{*}$of (\ref{eq:mp}) as in \cite{zhang2010icnp}.

As in Proposition \ref{prop:acyclic_property}, we label all the nodes with
indexes $\{I_{v}\}$ such that for any node $v\in R$, $I_{u}$ for
any $u\in in(v)$ is smaller than $I_{v}$. Note that under rate allocation
$\boldsymbol{r}^{*}$, the constraint (\ref{eq:mp_constraint1}) guarantees
that the aggregate incoming rate of each node is larger than or equal
to $z^{*}$. We show by induction that every node can receive innovative
contents at rate $z^{*}$ under our content scheduling policy.

We start with the node with index $1$. The only incoming neighbor
of this node is the source $s$, which generates new contents at rate
$z^{*}$. Thus the node with index $1$ receives innovative contents
at rate $z^{*}$. Suppose the nodes with index smaller than $i$ ($i>1$)
receive innovative contents at rate $z^{*}$. Note that these nodes
can be regarded as {}``\textit{sources}'' because new contents are
injected into each of them with the rate same as the source $s$.
Now we check the node with index $i+1$. Because the incoming neighbors
of this node are the subset of the {}``\textit{sources}'', and this
node manages to obtain as many innovative contents as possible from
its incoming neighbors, we can conclude that the node with index $i+1$
can receive innovative contents at rate $z^{*}$.

From Theorem \ref{thm:sufficient_necessary}, we know that the constraints
of (\ref{eq:mp}) are sufficient and necessary conditions for any
feasible broadcast rate. Thus we get $z^{*}=B$. So our broadcasting
algorithm is throughput optimal.
\end{IEEEproof}

\section{Simulation}

We implement our broadcasting algorithm using Python and conduct simulation
studies to evaluate the performance of our solution.

\subsection{Settings}

\begin{figure}
\begin{centering}
\includegraphics[scale=0.04]{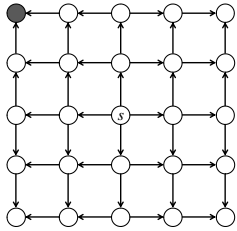}
\par\end{centering}
\caption{Topology of the simulated overlay network: The center node is the source. The
graph has a {}``grid'' topology. When the number of nodes increases,
the graph expands symmetrically as a 2D grid. The dark node is the broadcast bottleneck.}
\label{fig:grid_network}
\end{figure}

In our simulations, time is chopped into slots of equal length. The
topology of the overlay network is shown in Fig. \ref{fig:grid_network}.
We adopt two different settings. In Setting I, the capacity constraint
is on the edge. The edge capacity is $4$ Mbps except the incoming
edges of the top-left node (filled in black in Fig.
\ref{fig:grid_network}), at $1$ Mbps each. In Setting II, the capacity constraint is on the
nodes. The node capacity is $8$ Mbps except the source and
the incoming neighbors of the top-left node. The source's capacity
is $16$ Mbps. The capacity of each incoming neighbor of the top-left
node is $1$ Mbps. In both settings, given our capacity constraints,
the broadcast bottleneck is at the network edge and thus
far from the source. It's easy to check that the maximum broadcast
rate under both settings is $2$ Mbps. We choose the number of nodes
as $5^{2},15^{2},35^{2},105^{2}$ respectively. The utility function is $U(\cdot)=\log (\cdot)$.

Under Setting I and II, we next evaluate our broadcasting algorithm.

\subsection{Evaluation of the Proposed Broadcasting Algorithm}

We evaluate our broadcasting algorithm proposed in Section \ref{sub:broadcasting_algorithm}
and answer the following questions: 1) does it converge to the maximum
broadcast rate as expected by the theoretical analysis? 2) how fast
does it converge? 3) what's the impact of network size on the convergence
time (the length of the interval from the start to where the broadcast
rate begins to stay with the optimum)?

\begin{figure}
\subfigure[Broadcasting under Setting I\label{fig:broadcast_edge_constraint}]{\begin{centering}
\includegraphics[scale=0.20]{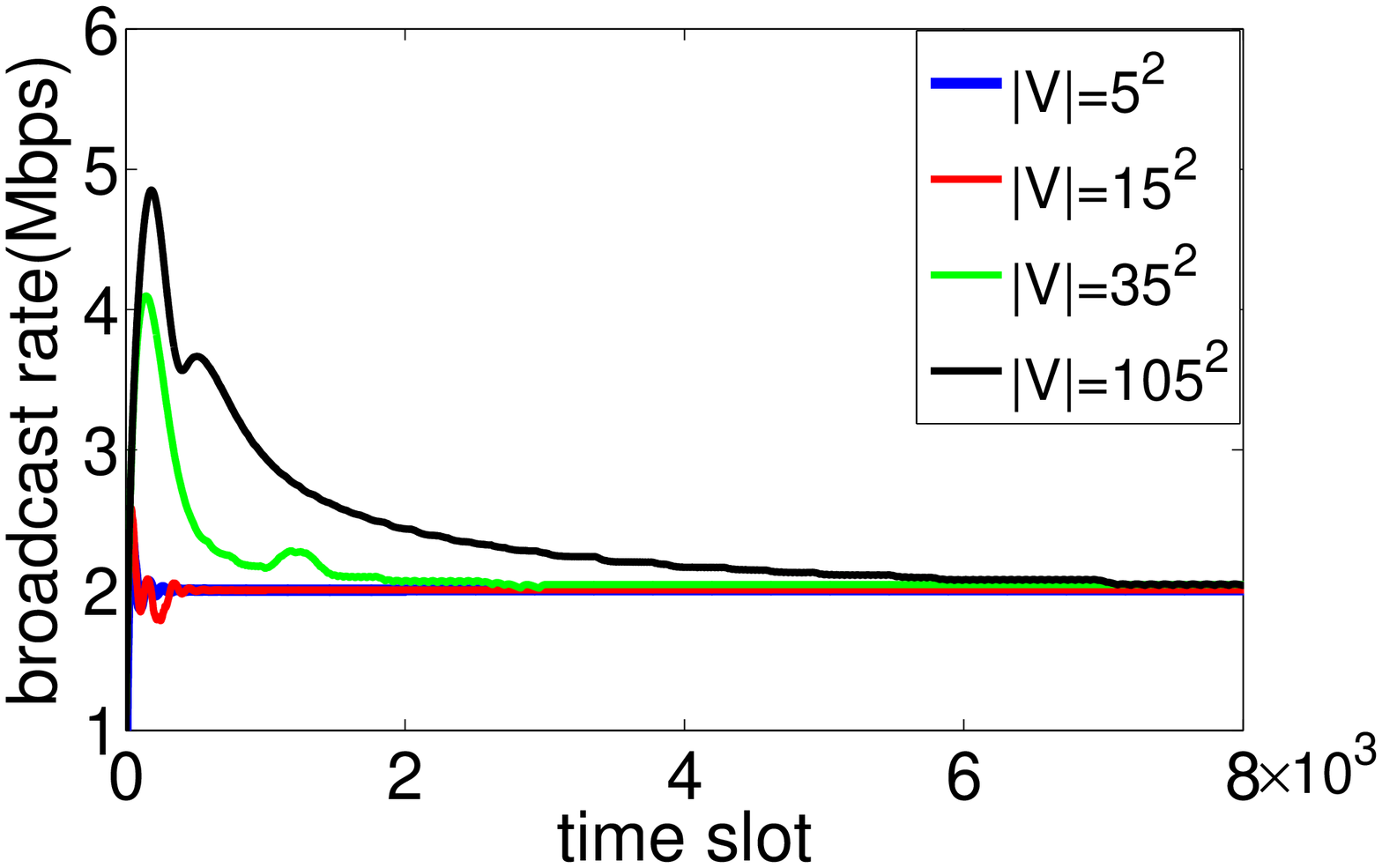}
\par\end{centering}
}\subfigure[Broadcasting under Setting II\label{fig:broadcast_node_constraint}]{\begin{centering}
\includegraphics[scale=0.21]{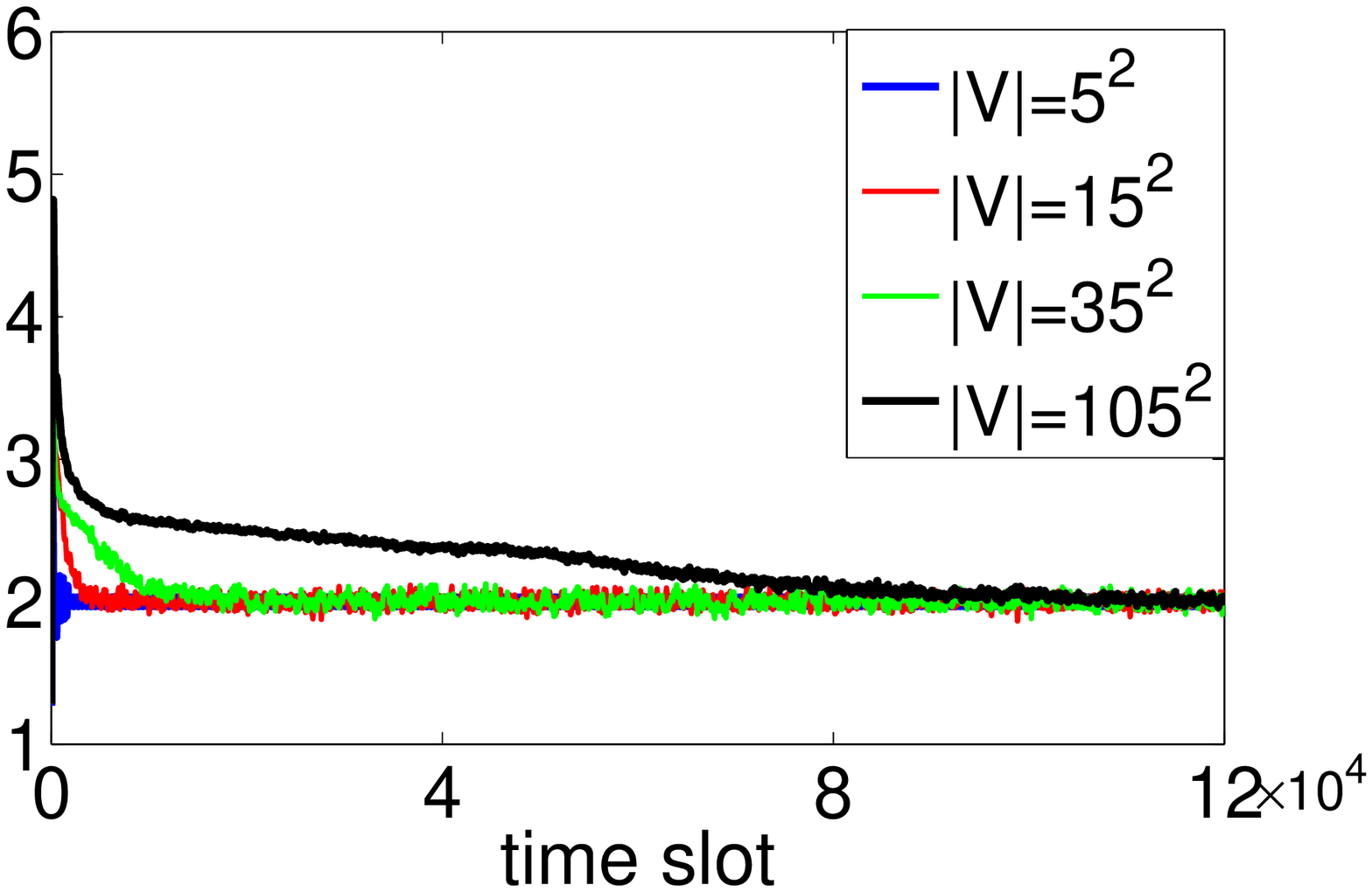}
\par\end{centering}
}
\caption{Simulation results}
\end{figure}

We show the results under Setting I in Fig. \ref{fig:broadcast_edge_constraint}
and the results under Setting II in Fig. \ref{fig:broadcast_node_constraint}.
We have the following observations. First, our broadcasting algorithm
can converge to the maximum broadcast rate $2$ Mbps and is thus optimal
under different network capacity models. Second, when the network
grows, the convergence time increases. This is because the bottleneck
(at the network edge) is further from the source as the number of
nodes gets larger. It takes longer for the source to learn about
the bottleneck and to adjust the broadcast rate correspondingly.
Third, under Setting I, the convergence times are about $300,600,2000,6000$ time slots respectively
as the number of nodes increases; under Setting II, the convergence times are
$1000,5000,19000,10^5$ time slots respectively.
We can see that the convergence time of our algorithm grows linearly with the
network size under both settings. That means our algorithm is suitable
to be implemented in large-scale systems. It might be an interesting
future direction to theoretically analyze the convergence behavior
of our algorithm. Forth, the convergence under Setting II is longer
than that under Setting I. The reason is that, different from Setting
I, each node needs to allocate its upload capacity among its neighbors
every time slot under Setting II.

\bibliographystyle{IEEEtran}
\bibliography{ref,p2p}

\end{document}